\documentclass{superfri}

\newcommand\dslash{\ensuremath{{\not}D}}

\begin{document}

\author{D.~Rohr\footnote{\label{fias}Frankfurt Institute for Advanced Studies, Goethe University Frankfurt, Department for High Performance Computing, Ruth-Moufang-Str. 1, 60438 Frankfurt, Germany, rohr@compeng.uni-frankfurt.de},
G.~Ne{\v s}kovi{\' c}\footnoteref{fias},
M.~Radtke\footnoteref{fias},
V.~Lindenstruth\footnoteref{fias}${}^,$\footnote{GSI Helmholtz Center for Heavy Ion Research, Planckstra�e 1, 64291 Darmstadt, Germany}}

\title{The L-CSC cluster: greenest supercomputer in the world in Green500 list of November 2014}

\maketitle{}

\begin{abstract}
The L-CSC (Lattice Computer for Scientific Computing) is a general purpose compute cluster built of commodity hardware installed at GSI.
Its main operational purpose is Lattice QCD (LQCD) calculations for physics simulations.
Quantum Chromo Dynamics (QCD) is the physical theory describing the strong force, one of the four known fundamental interactions in the universe.
L-CSC leverages a multi-GPU design accommodating the huge demand of LQCD for memory bandwidth.
In recent years, heterogeneous clusters with accelerators such as GPUs have become more and more powerful while supercomputers in general have shown enormous increases in power consumption making electricity costs and cooling a significant factor in the total cost of ownership.
Using mainly GPUs for processing, L-CSC is very power efficient, and its architecture was optimized to provide the greatest possible power efficiency.
This paper presents the cluster design as well as optimizations to improve the power efficiency.
It examines the power measurements performed for the Green500 list of the most power efficient supercomputers in the world which led to the number~1 position as the greenest supercomputer in November 2014.

\end{abstract}

\section*{Introduction}

Quantum Chromo Dynamics (QCD) is the physical theory of the strong force, which describes the interaction between quarks and gluons, the fundamental constituents of hadronic matter in the universe.
It is a highly nonlinear theory where perturbative methods are only applicable in a small regime.
Lattice QCD (LQCD) uses a discretization in a space time grid, and it is the only general a priory approach to QCD computations.
LQCD requires the inversion of the Dirac operator, which is usually performed by a conjugate gradient algorithm, which involves a sparse matrix-vector-multiplication called \dslash{}.
This \dslash{} operator is the computational hotspot of LQCD applications and it is responsible for a majority of the runtime of the program.
The bottleneck in \dslash{} is usually not the compute performance but the memory bandwidth, because sparse matrix-vector-multiplications require many memory loads per compute operation compared to other matrix operations with dense matrices like DGEMM.
Hence, for a compute cluster with LQCD as primary focus, a large memory bandwidth is paramount.

Supercomputers are inevitable in today's research.
Scientific challenges demand the fastest possible supercomputers, but it is prohibitively expensive to acquire more and more compute power through the use of more and more electricity.
In order to use the available resources to the maximum, computers have to become more power efficient.
During the last several years, heterogeneous HPC clusters which combine traditional processors with special accelerators such as GPUs or Xeon Phi have proven to deliver both superior compute performance and energy efficiency.
In an effort to raise awareness for power efficiency, the Green500 list~\cite{Sharma2006} provides a list of supercomputer power efficiencies and presents the ``greenest'' supercomputers in the world.

This paper presents L-CSC (Lattice Computer for Scientific Computing), which is built of commodity hardware and features four high-performance GPUs per compute node.
It is organized as follows:
Section~\ref{sec:cluster} describes the hardware of the cluster and why it is suited for LQCD.
The section outlines the design decisions for good power efficiency.
Section~\ref{sec:linpack} illustrates some optimizations we applied to achieve the best efficiency.
Finally, section~\ref{sec:green500} describes the efforts required to obtain an accurate and reasonable power measurement for the Green500 list and presents the results.

\section{The L-CSC cluster}
\label{sec:cluster}

In order to access a broad variety of hardware and to reduce acquisition costs, L-CSC is based on off-the-shelf components.
Its design follows the LOEWE-CSC and Sanam~\cite{bib:sanampaper} clusters, which have proven the validity of the commodity hardware approach for GPU accelerated HPC clusters.
L-CSC is a general purpose cluster that can run any kind of software, although its main focus is LQCD.

\looseness=-1
The most important criteria for a supercomputer with LQCD focus are memory bandwidth and memory capacity.
Memory bandwidth defines the compute performance and memory capacity defines the maximum lattice size.
The performance of \dslash{} depends more or less linearly on the memory bandwidth and it is possible to use a large fraction of the theoretically available bandwidth in the application (Bach et al.~\cite{pos} show above 100~GFLOPS which translates to about 80\% of the peak memory bandwidth with the OpenCL application employed on L-CSC).
The demands with respect to memory capacity are a bit more complex.
It is mandatory that the lattice fits in GPU memory.
If it fits, any additional memory cannot be used at all.
Hence, memory should not be chosen too large in the first place.
For L-QCD calculations, the extent of the time dimension of the lattice is anti-proportional to the temperature.
Thermal lattices ($T > 0$) need much less memory than lattices with~$T \approx 0$.
As a different aspect, the distance of the lattice points can be decreased for better accuracy requiring more memory, but this also slows down the program.
Hence, the answer to the question of how much memory is needed depends on the actual problem.
A memory of~$3$ GB is already enough for most thermal lattices sizes~($T > 0$)~\cite{bib:sanampaper}, but has some limitations.
By and large, we consider~$16$~GB of L-CSC's S9150 cards sufficient for almost all lattices.

To make things even more complex, one can distribute the lattice over multiple GPUs or even over different compute nodes.
Tests on the Sanam cluster have shown a performance decrease on the order of~20\%, when more than one GPU is used.
The paradigm for L-CSC is to run most lattices on a single-GPU only, while there is still the possibility of using multiple GPUs for very large ones.
Still, multiple GPUs inside a compute node can be fully used in parallel to compute independent lattices.
Since LQCD needs a lot of statistic involving a great deal of lattices, this approach is very efficient.

Overall, the design goal was four GPU boards per node with maximum aggregate GPU memory bandwidth - under the constraint of sufficient memory per GPU.
Two GPU types have been chosen: The AMD FirePro S9150 GPU, featuring a capacity of~$16$~GB and a bandwidth of~$320$~GB/s.
And the AMD FirePro S10000 dual-GPU (i.e. eight GPU chips per node), with a capacity of~$2 \times 6$~GB ($6$~GB per GPU chip) and a bandwidth of~$2 \times 240$~GB/s, thus with a higher aggregate bandwidth than~S9150.
Besides the higher memory capacity, the S9150 has the additional advantage of being able to reduce the wall time for small jobs compared to the S10000 due to the higher per-GPU-chip bandwidth.
This is important for testing when a quick answer is needed.
L-CSC runs all larger lattices on the S9150, and the smaller latices on both S10000 and S9150.
Very large lattices can span multiple S9150 cards, having access to~$64$~GB of GPU memory per node.

The L-CSC consists of 160 compute nodes with in total 48~S10000 GPUs and 592~S9150 GPUs.
Each compute node consists of an ASUS ESC4000 G2S/FDR server, two Intel \mbox{Ivy-Bridge-EP} ten-core CPUs, and 256~GB of DDR3-1600 memory.
In order to offer more flexibility for general purpose applications on the CPUs in parallel, two CPU models are used: 60 nodes have 3~GHz CPUs for applications with high CPU demands and 90 nodes have 2.2~GHz CPUs.
The interconnect is 56~GBit FDR InfiniBand with half bisectional bandwidth and fat-tree topology.
The OpenCL LQCD application achieves around~$135$~GFLOPS per S9150 GPU in \dslash{}.

\section{Optimizing for best power efficiency}
\label{sec:linpack}

The Linpack benchmark is the standard benchmark for measuring the performance of supercomputers.
The Green500 list presents the most power efficient supercomputers in the world~\cite{Sharma2006}.
Its ranking is determined by the GFLOPS achieved in the Linpack normalized by the average electricity consumption during the Linpack run.

Even though L-CSC consists of commodity hardware, there are no unnecessary components that drain power.
The main contributors are the CPUs, GPUs, memory, chipset, network, and remote management.
Power consumption of the hard disk with scratch space in each node and of other components are comparatively small, given that each node features four GPUs with~$275$~W each.
Universal Serial Bus (USB) contributes significantly with up to~$20$~W.
L-CSC uses full USB suspend which amounts to the same savings as if USB were switched of completely, so USB does not play a role here.

Some additional optimizations boost L-CSC's power efficiency during the Linpack run for the Green500.
An InfiniBand-based network boot allows switching off hard disks, sata controller, and all Ethernet LAN ports completely.
Optimizing the FAN settings by using a curve that defines different FAN duty cycles for different load levels ensures that the FANs always run only at the minimum speed required.
This reduces power consumption.

In addition, the version of the Linpack benchmark, HPL-GPU~\cite{bib:isc}, which we have developed and used for the LOEWE-CSC and Sanam clusters provides two operating modes.
One optimized for maximum performance, and an alternative mode that sacrifices a small fraction of the performance to reduce the power consumption resulting in better net power efficiency.
This alternative efficiency-optimized mode was developed further for L-CSC.
Now it offers the possibility of dynamically offloading some Linpack tasks usually executed on the processor onto the GPU, resulting in better power efficiency because of the increased efficiency of the GPU.

\section{Measuring the power consumption for the Green500 list}
\label{sec:green500}

The Green500 ranking is determined by the quotient of the achieved performance in the Linpack benchmark divided by the average power consumption during Linpack execution.
Due to late installation of the system, only~56~nodes with S9150 GPUs were available for the Linpack benchmark in November 2014, which were connected by three InfiniBand switches in a ring-configuration.
The measurement methodology is defined in~\cite{bib:eehpc}.
Tab.~\ref{levels} lists three measurement levels defined in the methodology document yielding different accuracies.

\tab{levels}{Measurement levels for Green500 with different accuracy}{
\begin{tabular}{llll}
\hline
Level & Components & Measured fraction of system & Duration \\
\hline
1 & Only compute nodes & At least~$\frac{1}{64}$ of the system & At least~20\% of the\\
& & & middle~80\% of the run \\
2 / 3 & Full cluster with network & At least~$\frac{1}{8}$ / full system & Full runtime \\
\hline
\end{tabular}
}

\looseness=-1
Level~1 is provided for facilities without sufficient equipment for higher level measurements.
Unfortunately, the level~1 specifications are exploitable such that one can create measurements which show a higher power efficiency than actually achieved.
Thus, higher levels are preferred.
The L-CSC installation had only one revenue grade powermeter~\cite{bib:eehpc} available, and it was thus impossible to measure a larger fraction of the system.
Thus, only a level~1 measurement was feasible.
All measures were taken to make the result as accurate as possible.
A level~1 measurement does not need to cover the entire duration~(see Tab.~\ref{levels}).
Toward the end of the run, the power consumption decreases steadily.
Hence, one can obtain a lower average power measurement by measuring between 70\% and 90\% of the runtime.
The L-CSC measurement covers the full duration.
Due to the lack of more revenue grade power measurement equipment, only two compute nodes could be measured.
Power consumption variability of nodes  can be estimated by measuring the efficiency of several individual nodes during single-node Linpack runs, which yielded the following values:
5154.1, 5260.1, 5248.4, 5245.5, 5125.1, 5301.2, 5169.3~[MFLOPS/W].
The results show a relatively small variation of $\pm 1.2$\% and the submitted measurement uses nodes with middle power consumption.
Finally, the measurement includes the power consumption of the InfiniBand switches, which was~$257$~W.
Hence, the difference to the full level three measurement is small, lying only in the smaller part that was measured, with an estimated inaccuracy of below~1.2\%.

The 56 nodes used for the measurement achieved a Linpack performance of~301.5~TFLOPS expensing on average 74.4~kW and yielding an average efficiency of~5271.8~MFLOPS/W.
With this result, L-CSC was awarded 1st place in the Green500 list of November 2014 as the most power efficient supercomputer in the world.

\ack{We would like to thank Advanced Micro Devices, Inc. (AMD) and ASUSTeK Computer Inc. (Asus) for their support. The work was partially funded by HIC for FAIR.}

\end{document}